\documentclass[ste,twoside]{stefano}

\usepackage{amsmath}
\usepackage{amssymb}
\usepackage{graphics}
\usepackage{rotating}
\usepackage{cite}
\usepackage{color}

\def\makeheadbox{{%
\hbox to0pt{\vbox{\baselineskip=10dd\hrule\hbox
to\hsize{\vrule\kern3pt\vbox{\kern3pt \hbox{  {\sc
quant-ph/0401145} } \hbox{  {\sc Phys. Rev. A {\bf 70}, 022101-11
(2004) } \hspace*{7.8cm} {\color{blue}{$\boldsymbol{\Sigma \delta
\Lambda}$}} }
\kern3pt}\hfil\kern3pt\vrule}\hrule}%
\hss}}}

\def\0{\mbox{\tiny $0$}}
\def\1{\mbox{\tiny $1$}}
\def\2{\mbox{\tiny $2$}}
\def\3{\mbox{\tiny $3$}}
\def\4{\mbox{\tiny $4$}}
\def\5{\mbox{\tiny $5$}}
\def\6{\mbox{\tiny $6$}}
\def\7{\mbox{\tiny $7$}}
\def\8{\mbox{\tiny $8$}}
\def\9{\mbox{\tiny $9$}}
\def\l{\mbox{\tiny $l$}}
\def\m{\mbox{\tiny $m$}}
\def\infm{\mbox{\tiny $-\infty$}}
\def\infp{\mbox{\tiny $+\infty$}}
\def\E{\mbox{\tiny $E$}}
\def\Ep{\mbox{\tiny $E'$}}
\def\mi{\mbox{\tiny $-$}}
\def\pl{\mbox{\tiny $+$}}
\def\ppm{\mbox{\tiny $\pm$}}

\begin{document}
%

\title{\Large SURVIVAL LAW IN A POTENTIAL MODEL}

\author{
Stefano De Leo\inst{1}
\and Pietro P. Rotelli\inst{2} }

\institute{
Department of Applied Mathematics, State University of Campinas\\
PO Box 6065, SP 13083-970, Campinas, Brazil\\
{\em deleo@ime.unicamp.br}
\and
Department of Physics, INFN, University of Lecce\\
PO Box 193, 73100, Lecce, Italy\\
{\em rotelli@le.infn.it}
}


\date{{\em January, 2004} - Revised version:  {\em April, 2004} }

\abstract{The radial equation of a simple potential model has long
been known to yield an exponential survival law in lowest order
(Breit-Wigner) approximation. We demonstrate that if  the
calculation is extended to fourth order the survival law exhibits
a parabolic short time behavior which leads to the quantum Zeno
effect. This model has further been studied numerically to
characterize the extra exponential time parameter which
compliments the lifetime. We also investigate the inverse Zeno
effect.}










\titlerunning{\sc the quantum zeno effect in a potential model}

\maketitle


\section*{I. INTRODUCTION}

The exponential survival law is known to be an excellent
phenomenological fit to unstable phenomena. However, there is no
rigorous derivation of this law in quantum mechanics. Most text
book derivations are "classical" in nature since they refer only
to decay probabilities. In quantum mechanics, one would require an
exponential time dependence for the survival amplitude \cite{CT}.
To achieve this one must make approximations. Standard procedure
is to consider a tunnelling process, as we do in this paper. Some
earlier important theoretical papers upon tunnelling are given in
Refs. \cite{NREF}.

Quantum mechanics allows us to say that the survival probability
$P(t)$ is definitely {\em not} exponential for short and long
times. For short times it can be argued that a power expansion in
$t$ must lack the linear term, i.e.
$(\mbox{d}P/\mbox{d}t)_{\mbox{\tiny
$t\hspace*{-0.08cm}=\hspace*{-0.06cm}0$}}=0$. This result when
combined with the hypothesis of "frequent" measurements leads to
what is known as the quantum Zeno effect (QZE) \cite{QZE}. It was
first named the quantum Zeno paradox by Misra and
Sudarshan\cite{MS77} precisely because it was considered a false
result\cite{noqze}. Nowadays, the QZE is generally acknowledged as
a real phenomenon and indeed there are a number of experiments
which claim to have verified it and others are planned\cite{IHBW}.
Once accepted, it is even possible to predict a QZE within a
classical calculation\cite{timelapse}. As for the long time
behavior, this is predicted to be a power law behavior in $t$ .
This latter result can be derived even for a Breit-Wigner spectrum
by imposing a low energy cut-off\cite{longt} (which is expected on
physical grounds). The long-time property of $P(t)$ will not be
treated specifically in this paper but it is referred to in the
concluding Sections.

Returning to the short time behavior, the simplest demonstration
of the non-exponential behavior of the survival law is based upon
the hermitian nature of the Hamiltonian\cite{NNP} (when all
channels are considered). Consider a state $|\psi(t) \rangle$
initially in a (quasi-bound) state represented by $|\psi(0)
\rangle=|\psi_{\0} \rangle$. It is assumed that $|\psi_{\0}
\rangle$ can "decay" into another or other states. At time $t$ the
state will have evolved into
\[ |\psi(t) \rangle= \exp( - i H t ) \, |\psi_{\0} \rangle \, \, ,\]
where $H$ is the Hamiltonian. Expanding in a power series of $t$,
\begin{eqnarray*}
|\psi(t) \rangle & = & \left[ \, 1 - i H t - \mbox{$\frac{1}{2}$}
\, H^{\2} t^{\2} + \mbox{O}(t^{\3}) \, \right] \, |\psi_{\0}
\rangle \, \, .
 \end{eqnarray*}
  The amplitude for non-decay is given, up to a possible
 phase, by
\begin{eqnarray*}
\langle \psi_{\0} | \psi(t) \rangle  & = &  1 - i \, \langle
\psi_{\0} | H | \psi_{\0}\rangle \, - \mbox{$\frac{1}{2}$} \,
\langle \psi_{\0} | H^{\2} | \psi_{\0}
\rangle \, t^{\2} + \mbox{O}(t^{\3}) \nonumber \\
 & \equiv &
1 - i \, \langle H \rangle_{\0} \, t - \mbox{$\frac{1}{2}$} \,
\langle H^{\2} \rangle_{\0} \, t^{\2} + \mbox{O}(t^{\3})\, \, ,
\end{eqnarray*}
where $\langle \, \, \rangle_{\0}$ stands for the average over the
state $| \psi_{\0} \rangle$. Whence, the non-decay probability is,
\begin{eqnarray*}
P(t)  & \equiv  & |\langle \psi_{\0} | \psi(t) \rangle|^{\2} \\
      & = &   1 - i \, \langle \left( H - H^{\pl} \right) \rangle_{\0}
      - \mbox{$\frac{1}{2}$} \, \left( \langle H^{\2}
     \rangle_{\0} +  \langle H^{\pl \, \2}
     \rangle_{\0}  - 2 \, \langle  H \rangle_{\0}  \langle H^{\pl}
     \rangle_{\0} \right) \, t^{\2}  + \mbox{O}(t^{\3}) \, \, .
\end{eqnarray*}
Now, using $H=H^{\pl}$, we find
\[
P(t) = 1 - t^{\2} \left( \langle H^{\2}
     \rangle_{\0} - \langle H \rangle_{\0} ^{\, \2}
\right)    + \mbox{O}(t^{\3}) \, \, .
     \]
The linear term in $t$ which corresponds to the linear term in $H$
has cancelled. Of course this derivation assumes the existence of
$\langle H^{\2} \rangle_{\0}$  , in addition to $\langle H
\rangle_{\0}  $. It is useful to observe at this point that if
this demonstration where extended to all powers in $t$ then not
only would the linear term vanish but {\em all } odd powers of $t$
would vanish. We shall return to this when we discuss some
numerical calculations in Section IV.

The QZE merits a name (even if this choice is not quite
appropriate) because it implies a potentially spectacular
phenomenon - the prediction that "frequent" short time tests of
the state of an unstable system will inhibit its decay. Much has
been written upon this, in particular upon the question of what
may constitute an "observation" or test of the system. We wish to
add here only a comment upon this fascinating subject. A
measurement of a system normally produces a collapse of the wave
function. It is therefore perfectly reasonable to expect that
observations should modify, for example, the survival law.
However, the exponential curve $e^{-t / \tau}$ is unique in this
respect because it is not altered by measurements. This fact is
connected to the mathematical feature that the average value
$\langle (t-t_{\0}) \rangle$ from $t_{\0}$ to $\infty$ (lifetime)
is independent of the lower limit. This is the reason one does not
need to know when the, otherwise identical, unstable particles in
a "sample" where created in order to measure their lifetime. Even
if each had been created at a different time, one can treat them
as if "newly created" at the conventional time $t_{\0}$. In more
colorful terms, a series of operations of "cut and paste"
(measurements) are undetectable only if performed upon a single
exponential function. The QZE is thus a consequence of a
particular example of the non exponential nature of $P(t)$. What
is somewhat peculiar is that the absence of the linear term in $t$
is more important (because in principle more easily verifiable)
than the absence of any other or indeed of all the other odd
powers of $t$.

In this paper we wish to examine the survival law with the aid of
a particular potential model (Section II), already used in the
literature for a derivation of the exponential law, albeit as an
approximation\cite{potmod}. In Section III, we will repeat this
derivation with particular emphasis upon the
assumptions/approximations that lead to the exponential law. We
will then go beyond the lowest order calculation (Breit-Wigner
form) to a two pole approximation. Recently a two pole
approximation in a quantum field model was introduced by Facchi
and Pascazio\cite{FP2002}. In Section IV, we shall show that at
this more sophisticated level we predict the necessary conditions
for the QZE. Indeed, we speculate in the conclusions that only the
lowest order approximation loses this effect. In Section V, we
present the results of some numerical calculations and in
particular the dependence of an extra time constant upon the
parameters of the model. We will phenomenologically parameterize
these dependencies. We conclude in Section VI with the resume of
our results, a discussion of the "inverse Zeno effect" \cite{ize}
(described in Section V) and some observations upon the
significance of the extra time parameters.

\section*{II. THE POTENTIAL MODEL}

The starting point of our analysis is the three dimensional
Schr\"odinger equation for a particle of mass $m$ in a spherically
symmetric potential $V(r)$, i.e. a function only of the magnitude
$r$ of $\boldsymbol{r}$. Explicitly, we will use

\begin{picture}(180,90) \thinlines
\put(2,10){\vector(0,1){68}} \put(0,80){$V(r)$}
\put(2,10){\vector(1,0){148}} \put(153,10){$r$}
\put(7,32){\mbox{\small \sc Region 1}} \put(56,32){\mbox{\small
\sc Region 2}} \put(105,32){\mbox{\small \sc Region 3}}
\put(49,0){$a$} \put(98,0){$b$} \thicklines
\put(50,10){\line(0,1){35}} \put(50,45){\line(1,0){50}}
\put(100,45){\line(0,-1){35}}
\put(2,21){.....................................................}
\put(-12,42){$V_{\0}$} \put(-1,45){\line(1,0){3}}
\put(-12,20){$E$} \put(-1,22){\line(1,0){3}} \put(230,42){$ V(r) =
\left\{
\begin{array}{lcl}
0 &  & 0 < r < a \, ,\\
V_{\0} & ~~~~~ & a <  r <  b \, ,\\
0 &  & b < r \, .
\end{array}
\right.$}
\end{picture}

\noindent While our subject matter is not a stationary state
problem, it can be conveniently analyzed with the use of the
stationary state solutions, each of which has a simple (phase)
time dependence. Due to the spherical symmetry of the potential
energy and its time-independence, the Schr\"odinger equation
\begin{equation}
\label{seq} \left[ - \, \frac{\nabla^{\2}}{2\,m}  + V(r) \right]
\, \psi (\boldsymbol{r},t) = i \, \frac{\partial}{\partial t} \,
\psi (\boldsymbol{r},t)\, \, ,
\end{equation}
can be separated by using the energy eigenstates
\[
\psi (\boldsymbol{r},t) = R_{\E}(r) \, Y_{\l}^{\m}(\theta,\phi) \,
\exp(-iEt)  \, \, ,
\]
where $Y_{\l}^{\m}(\theta,\phi)$ are the spherical harmonics. Thus
Eq.(\ref{seq}) yields the well known time-independent equation
\begin{equation}
\label{qse2} \left[  - \, \frac{1}{2\,m} \left(
\frac{\partial^{\2}}{\partial r^{\2}} + \frac{2}{r} \,
\frac{\partial}{\partial r} \right) + \frac{l(l+1)}{2\,m\,r^{\2}}
+ V(r) \right] \, R_{\E}(r) \, Y_{\l}^{\m}(\theta,\phi)  = E \,
R_{\E}(r) \, Y_{\l}^{\m}(\theta,\phi) \,  \, .
\end{equation}
For a given value of $E$ and angular momentum, there are two
linearly independent solutions of the above second order equation,
which at the origin go like $r^{\l}$ or $1/r^{\mbox{\tiny
$l+1$}}$. However, those which behave like $1/r^{\mbox{\tiny
$l+1$}}$ must be rejected since it can be shown that
$Y_{\l}^{\m}(\theta,\phi)/r^{\mbox{\tiny $l+1$}}$ is not a
solution of the above eigenvalue equation for $r=0$. This is
because the Laplacian of $Y_{\l}^{\m}(\theta,\phi)/r^{\mbox{\tiny
$l+1$}}$ involves the $l$-th derivative of the Dirac delta
function $\delta(\boldsymbol{r})$ \cite{CT}.

Henceforth, we shall limit our attention to $s$-wave solutions
($l=0$) and consequently, the previous equation reduces to
\begin{equation}
\label{qse3} \left[ - \, \frac{1}{2\,m} \left(
\frac{\partial^{\2}}{\partial r^{\2}} + \frac{2}{r} \,
\frac{\partial}{\partial r} \right) + V(r) \right] \, R_{\E}(r)  =
E \, R_{\E}(r) \, \, ,
\end{equation}
with $R_{\E}(0)$ which, in accordance with our previous
discussion, is a constant. If we now put
\[
R_{\E}(r) = u_{\E}(r) / r \, \, ,
\]
we obtain the equation for the modified radial wave function
$u_{\E}(r)$
\begin{equation}
\label{qseu}  u_{\E}''(r) =  2\, m \,[ \, V(r) - E \,] \,
u_{\E}(r) \, \, ,
\end{equation}
where the prime stands for $d/dr$. The acceptable solution of
Eq.(\ref{qseu}) must go to zero at the origin, $u_{\E}(0)=0$. The
standard procedure for finding the stationary solutions can now be
applied. The general solution for $u_{\E}(r)$ and {\em any}
$E<V_{\0}$ in the three regions are
\begin{equation}
 u_{\E}(r) \, = \,
\left\{
\begin{array}{lclclcl}
 \mbox{\small \sc Region 1:} &~~~~ & 0 < r < a \, ,& ~~~ & C_{\1} \sin(kr)
 & &
 ~~~[\, k=\sqrt{2\,mE} \, \,]\, , \\
\mbox{\small \sc Region 2:} & &a < r < b \, ,&  & C_{\2} \, e^{\,
\rho r} + D_{\2} \,  e^{-\rho r}  & &
~~~[\, \rho=\sqrt{2\,m(V_{\0} - E)} \, \,] \, ,\\
\mbox{\small \sc Region 3:} &   & b < r \, ,&  & C_{\3} \, e^{i k
r} + D_{\3} \, e^{- i k r} \, .  &  &
\end{array}
\right.
\end{equation}
The requirement of continuity of $u_{\E}$ and $u'_{\E}$ at $r=a,b$
(a direct consequence of the differential equation for $u_{\E}$)
can be conveniently expressed in matrix form as
\[
\left[ \begin{array}{r} C_{\1} \sin(ka) \\ k \, C_{\1} \cos(ka)
\end{array}
\right] = \left( \begin{array}{rr} e^{\, \rho a} & e^{-\rho a} \\
\rho \, e^{\, \rho a} &  \, \, - \rho \, e^{-\rho a} \end{array}
\right) \, \left[
\begin{array}{r} C_{\2} \\ D_{\2} \end{array}  \right]~,
\]
and
\[
\left( \begin{array}{rr} e^{ikb} &  e^{-ikb}\\
i k \, e^{ikb}& \, \, - ik  \, e^{-ikb}  \end{array} \right)
\,\left[
\begin{array}{l} C_{\3} \\ D_{\3}
\end{array}
\right] = \left( \begin{array}{rr} e^{\, \rho b} & e^{-\rho b} \\
\rho \, e^{\, \rho b} & \, \,- \rho \, e^{-\rho b} \end{array}
\right) \, \left[
\begin{array}{r} C_{\2} \\ D_{\2} \end{array}  \right]~.
\]
Eliminating $C_{\2}$ and $D_{\2}$, we obtain $C_{\3}$ and $D_{\3}$
in terms of $C_{\1}$,
\begin{equation}
\label{cond} \left[
\begin{array}{l} C_{\3} \\ D_{\3}
\end{array}
\right] =\left( \begin{array}{rr} e^{-ikb} &\, \, \,
-\, \mbox{$\frac{i}{k}$} \, e^{-ikb}\\
 e^{ikb}& \, \, \mbox{$\frac{i}{k}$}  \, e^{ikb} \end{array}
 \right)\,
 \left( \begin{array}{rr} e^{\, \rho b} & e^{-\rho b} \\
\rho \, e^{\, \rho b} & \, \,- \rho \, e^{-\rho b} \end{array}
\right) \,\left( \begin{array}{rr} e^{-\rho a} & \, \,
\mbox{$\frac{1}{\rho}$}
\, e^{-\rho a} \\
 e^{\, \rho a} &  \, \, \, - \, \mbox{$\frac{1}{\rho}$} \, e^{\, \rho a}
\end{array} \right) \, \left[ \begin{array}{r}  \sin(ka) \\ k \, \cos(ka)
\end{array}
\right] \, C_{\1} / 4~.
\end{equation}
>From this equation it is straightforward to show that
$|C_{\3}|=|D_{\3}|$. The solutions in region 3 involve waves
"moving" in both directions.

The normalization of the $u_{\E}(r)$ is determined by the standard
condition for a continuous spectrum
\begin{equation}
\label{norm}
\int_{\0}^{\infp} \hspace*{-.4cm} \mbox{d}r \, \,
u_{\E}(r) \, u_{\Ep}^*(r)
 = \, \delta ( E-E')\, \, .
\end{equation}
We recall that  the definition of $R_{\E}(r)=u_{\E}(r)/r$
eliminates the factor $r^{\2}$ in the integration variables
($\mbox{d}^{\3}r=r^{\2}\,\mbox{d}r\,\mbox{d}\Omega$)
 while the $4\pi$ from the angular integration is accounted for in
 the definition of the spherical harmonic $Y_{\0}^{\0}$. The above normalization
 condition is dominated by region 3 since it has infinite range and
 determines uniquely $|C_{\3}|(=|D_{\3}|)$
 \[
 \left( \, |C_{\3}|^{\2} + |D_{\3}|^{\2} \, \right)
\int_{\0}^{\infp} \hspace*{-.4cm}\mbox{d}r \, \, \exp
[\,i\,(k-k')\,r\,] = \, m \, \delta ( k-k') \, / k
~~~~~~~~~~(k,k'\ge 0)\, \, ,
\]
whence
\begin{equation}
\label{c3}
 |C_{\3}|^{\2} = |D_{\3}|^{\2}= \, m \, / \, (2 \,\pi k)\,
\, .
\end{equation}
With this condition we can calculate all the other coefficient
moduli. We shall need in what follows the explicit value of
$|C_{\1}|^{\2}$. From Eq.(\ref{cond}), after some manipulations,
we find
\begin{equation}
\label{rel}
 |C_{\3}|= \mbox{$\frac{1}{4}$} \, \left| \, \exp [\,
\rho \,(b-a) \, ] \, \alpha_{\pl} \left( 1 - i \,
\mbox{$\frac{\rho}{k}$} \right) + \, \exp [ - \rho \, (b-a)\, ] \,
\alpha_{\mi} \left( 1 + i \, \mbox{$\frac{\rho}{k}$} \right) \,
\right| \, |C_{\1}|
\end{equation}
where,
\[
\alpha_{\ppm} = \sin(ka) \pm \, \mbox{$\frac{k}{\rho}$} \,
\cos(ka)~.
\]
Whence, by using Eqs.\, (\ref{c3}) and (\ref{rel})
\begin{equation}
\label{c1} |C_{\1}|^{\2} = \mbox{$\frac{8mk}{\pi}$}
 \, \left\{ \,
\left[ \, \exp (2\, a \rho\, w  )  \, \alpha_{\pl}^{\2} + \,
\exp(- \, 2 \, a  \rho \, w ) \, \alpha_{\mi}^{\2} \, \right] \,
(k^{\2}+ \rho^{\2}) + \,2 \, \alpha_{\pl} \alpha_{\mi} \,(k^{\2}-
\rho^{\2})\, \right\}^{- \1} \, \, ,
\end{equation}
where
\[ w = (b-a)/a\,\,.\]
 Now,  $\exp ( 2\, a \rho\, w  )$
  can be considered to be {\em a very large number
for the purposes of this study}. This can always be guaranteed for
$E<V_{\0}$ by increasing the width $b-a$ of the potential. Under
these conditions, in general
 the energy eigenstates have small values of $|C_{\1}|$
because of the large value of the first term in the square
brackets above. There exist however, quasi-stationary states
defined by $\alpha_{\pl}=0$ when the opposite is true. These
quasi-stationary energy levels coincide with the discrete spectrum
of an associated potential model,

\begin{picture}(180,90) \thinlines
\put(2,10){\vector(0,1){68}} \put(0,80){$V(r)$}
\put(2,10){\vector(1,0){148}} \put(153,10){$r$}
\put(7,32){\mbox{\small \sc Region 1}} \put(80,32){\mbox{\small
\sc Region 2}}  \put(49,0){$a$} \thicklines
\put(50,10){\line(0,1){35}} \put(50,45){\line(1,0){100}}
\put(2,21){.....................................................}
\put(-12,42){$V_{\0}$} \put(-1,45){\line(1,0){3}}
\put(-12.5,20){$E_{\0}$} \put(-1,22){\line(1,0){3}} \put(230,42){$
V(r) = \left\{
\begin{array}{lcl}
0 &  & 0 < r < a \, ,\\
V_{\0} & ~~~~~ & a <  r \, \, .
\end{array}
\right.$}
\end{picture}

\noindent The energy eigenvalues of this auxiliary potential,
indicated generically by $E_{\0}$ in the following, are given by
the solutions of the transcendental equation $\alpha_{\pl}=0$,
i.e.
\begin{equation}
\label{tran}
 \tan (ak_{\0}) = - \, k_{\0} / \rho_{\0}\, \, ,
\end{equation}
where $k_{\0}=\sqrt{2\,m E_{\0}}$ and $\rho_{\0}=\sqrt{2\,m
(V_{\0} - E_{\0})}$. The explicit values for a given $V_{\0}$ and
well size $a$ can be calculated numerically when needed. We shall
name the eigenfunctions of this second potential  $u_{\0}(r)$
\begin{equation}
u_{\0}(r) \, = \, \left\{
\begin{array}{lclcl}
 \mbox{\small \sc Region 1:} &~~~~ & 0 < r < a \, ,& ~~~ & c_{\1}
 \sin(k_{\0}r)\, ,\\
\mbox{\small \sc Region 2:} & &a < r  \, ,&  & d_{\2}  \exp (-
\rho_{\0} r ) \, .
\end{array}
\right.
\end{equation}
The continuity conditions for $u_{\0}(r)$ and  $u'_{\0}(r)$ yield
both  the transcendental equation (\ref{tran}) and
\begin{equation}
c_{\1} = \pm \, d_{\2} \, \exp (- \rho_{\0} a )\,
\sqrt{k_{\0}^{\2}+\rho_{\0}^{\2}} \, / \, k_{\0}\, \, .
\end{equation}
Finally, from the normalization of $u_{\0}(r)$ we obtain
\begin{equation}
\label{c10}
 |c_{\1}|^{\2} = 2 \,\rho_{\0}\,/\,(1 + a \rho_{\0}) \,
\, .
\end{equation}

\section*{III. THE BREIT-WIGNER APPROXIMATION}

In the previous section, we outlined the potential model and
introduced the variables and wave functions, including that of the
auxiliary potential with bound states $u_{\0}(r)$ whose energy
eigenvalues equal those of the quasi-stationary energies of our
potential. We begin here by assuming that initially our radial
state is $u_{\0}(r)$. Since this is not an eigenstate of our
potential but of the auxiliary potential, it will not remain
localized permanently but eventually leak into region 3 through
tunneling. To be more precise $u_{\0}(r)$ already has a small tail
outside the potential barrier, however this can be safely
neglected as we do in Eq.(\ref{co2}) below.

 We now decompose $u_{\0}(r,t=0)$ into energy eigenstates
$u_{\E}(r)$
\[
u_{\0}(r,t=0) \equiv u_{\0}(r) = \int_{\0}^{\infp} \hspace*{-.4cm}
\mbox{d}E \, \,  g_{\0}(E) \, u_{\E}(r)
\]
 from which it follows that
\begin{equation}
u_{\0}(r,t) = \int_{\0}^{\infp} \hspace*{-.4cm} \mbox{d}E \, \,
g_{\0}(E) \, u_{\E}(r) \, e^{-iEt} \, .
\end{equation}
 The spectrum $g_{\0}(E)$ is explicitly
derived in Eqs.\,(\ref{co2}) and ({\ref{co3}) below .
 At any time $t>0$, the amplitude for
finding the particle still in the quasi-stationary  state
$u_{\0}(r)$ is given by
\begin{equation}
\label{co}
 \int_{\0}^{\infp} \hspace*{-.4cm} \mbox{d}r \, \,
u_{\0}^*(r,0) \, u_{\0}(r,t)
 \,
 \, ,
 \end{equation}
and hence the non-decay or survival probability $P(t)$ is
\begin{eqnarray*}
P(t) & = & \left| \, \int_{\0}^{\infp} \hspace*{-.4cm}  \mbox{d}r
\, \,
u_{\0}^*(r,0) \, u_{\0}(r,t)  \, \right|^{\2}\\
 & = & \left| \,
\int_{\0}^{\infp} \hspace*{-.4cm} \mbox{d}E \, \, g_{\0}(E) \,
e^{-iEt} \, \int_{\0}^{\infp} \hspace*{-.4cm}  \mbox{d}E' \, \,
g_{\0}^*(E') \, \int_{\0}^{\infp} \hspace*{-.4cm} \mbox{d}r \, \,
u_{\Ep}^*(r) \, u_{\E}(r) \, \right|^{\2} \, \, ,
\end{eqnarray*}
which, after using the normalization condition Eq.(\ref{norm}),
becomes
\begin{equation}
\label{pge} P(t) = \left| \, \int_{\0}^{\infp} \hspace*{-.4cm}
\mbox{d}E \, \, |g_{\0}(E)|^{\2} \, e^{-iEt} \, \right|^{\2} \, \,
.
\end{equation}
The index upon $g_{\0}(E)$ reminds us that we have chosen one of
the possible many quasi-stationary states $E_{\0}$. To determine
$g_{\0}(E)$ we use
\[
\int_{\0}^{\infp} \hspace*{-.4cm} \mbox{d}r \, \, u_{\E}^*(r) \,
u_{\0}(r,0)
 = \int_{\0}^{\infp} \hspace*{-.4cm} \mbox{d}E' \, \, g_{\0}(E') \,
\int_{\0}^{\infp} \hspace*{-.4cm} \mbox{d}r \, \, u_{\E}^*(r) \,
u_{\Ep}(r) = g_{\0}(E)\, \, .
\]
The left hand side of this equation can easily be calculated if
one assumes $b \gg a$ so that $u_{\0}(r,0)$ can be considered
negligible for $r>b$. This is the first of the approximations
made. Now, as we shall see, for suitable choices of parameters,
$g_{\0}(E)$ is highly peaked around the quasi-energy $E_{\0}$
(second approximation compatible with the first). In these cases,
the function $u_{\E}(r)$ does not, practically, differ  from
$u_{\0}(r,0)$ in region 1 and  2 except for its normalization. We
may then approximate the above integral to an integral over only
region 1 and 2 and use
\[
u_{\E}^*(r) \approx C_{\1} \, u_{\0}^*(r,0) \, / \, c_{\1}
~~~~~~~~~~(r<b )\, \, ,
\]
whence for $E$ around $E_{\0}$
\begin{equation}
\label{co2}
 g_{\0}( E) \approx (C_{\1} / c_{\1} ) \,
\int_{\0}^{b} \hspace*{-.2cm} \mbox{d}r \, \, u_{\0}^*(r,0) \,
u_{\0}(r,0) \approx (C_{\1} / c_{\1} ) \,  \int_{\0}^{\infp}
\hspace*{-.2cm} \mbox{d}r \, \,  u_{\0}^*(r,0) \, u_{\0}(r,0)  =
C_{\1} / c_{\1}\, \, ,
\end{equation}
otherwise
\begin{equation}
\label{co3}
 g_{\0}( E) \approx 0 \, \, .
\end{equation}
By using Eqs.\,(\ref{c1}) and (\ref{c10}), we then obtain (for $E$
around $E_{\0}$)
\[
 |g_{\0}(E)|^{\2} \approx \mbox{$\frac{4m k\,(1+a \rho_{\0})}{\pi \rho_{\0}}$} \, \left\{ \,
\left[ \, \exp(2\, a \rho\, w) \, \alpha_{\pl}^{\2} + \, \exp (- 2
\, a\rho \, w) \, \alpha_{\mi}^{\2} \, \right] \, (k^{\2} +
\rho^{\2}) + \,2 \, \alpha_{\pl} \alpha_{\mi} \,(k^{\2} -
\rho^{\2}) \, \right\}^{- \1} \, \, .
\]
For convenience, we now make a change  of variables from $E$ and
$t$, to $\sigma$ and $\tilde{t}$ , by defining the dimensionless
quantities
\[
\sigma=ak ~~~~~\mbox{and}~~~~~\tilde{t}=t/(2\,ma^{\2})\, \, .
\]
Henceforth, we will indicate by $E_{\0}$ the {\em chosen} initial
quasi-stationary eigenvalue. In these new variables,
\begin{equation}
\label{int} P(t) = \left[ \, 4 (1+a \rho_{\0}) / a \pi \rho_{\0}
\, \right]^{\2} \, \left| \, \int_{\0}^{\infp} \hspace*{-.4cm}
\mbox{d}\sigma \, \left[ \, \exp (-i \sigma^{\2} \tilde{t}\,)\,  /
f(\sigma) \, \right] \, \right|^{\2} \, \, ,
\end{equation}
where
\begin{eqnarray}
f(\sigma) &= & \left\{ \, \left[ \, \exp \left( \, 2 \, \sqrt{2\,
ma^{\2}V_{\0}- \sigma^{\2}} \, w \right) \, \left( \, \sin \sigma
+ \, \frac{\sigma \, \cos \sigma}{\sqrt{2\, ma^{\2}V_{\0} -
\sigma^{\2}} } \,
\right)^{\2} \right. \right. + \nonumber \\
 & & \hspace*{.6cm} \left. \left.
\exp \left( \, - 2 \, \sqrt{2 \, m a^{\2}V_{\0} - \sigma^{\2}} \,
\, w \right) \, \left( \, \sin \sigma - \, \frac{\sigma \, \cos
\sigma}{\sqrt{2\, m a^{\2}V_{\0} - \sigma^{\2}} } \, \right)^{\2}
\, \, \right] \, \frac{2\,
m a^{\2}V_{\0} }{\sigma^{\2}}  \right. + \nonumber \\
 & & \hspace*{.4cm} \left. 2 \, \left( \, \sin^{\2} \sigma - \,
\frac{\sigma^{\2} \, \cos^{\2} \sigma}{2 \, m a^{\2}V_{\0} -
\sigma^{\2} } \, \right) \,\frac{2 \, (\sigma^{\2} -  m
a^{\2}V_{\0}) }{\sigma^{\2}}\, \right \} \, \, .
\end{eqnarray}
In accordance with our second approximation, we will assume that
$1/f(\sigma)$ is peaked about $\sigma_{\0}$ corresponding to the
quasi-stationary energy $E{\0}$ ($\sigma_{\0} =
a\sqrt{2\,mE_{\0}}$). We then expand $f(\sigma)$ around
$\sigma_{\0}$ and  keep only up to quadratic terms in $\Delta
\sigma =\sigma - \sigma_{\0}$. The validity of the peaked form is
not in question. It can eventually be verified a posteriori.
However, our assumption neglects terms of  order  $(\Delta
\sigma)^{\3}$ and higher. We shall indicate this fact by a
superscript $2$ on $f(\sigma)$,
\begin{equation}
\label{qbw} f^{\mbox{\tiny $[2]$}}(\sigma) = \exp( 2 \, a
\rho_{\0}\, w) \, \left[ \, (1+a \rho_{\0}) \, ( k_{\0}^{\2} +
\rho_{\0}^{\2}) \, \right]^{\2}  \, \left[ \, (\sigma -
\sigma_{\0})^{\2} - \epsilon \, (\sigma - \sigma_{\0}) +
\gamma^{\2} \, \right] /  a^{\2} k_{\0}^{\2}\, \rho_{\0}^{\4} \,
\, ,
\end{equation}
where
\[ \epsilon = \exp ( - \,2 \, a \rho_{\0} \, w) \, \frac{4\,ak_{\0}\rho_{\0}^{\2
} \, (k_{\0}^{\2} - \rho_{\0}^{\2})}{(1+a \rho_{\0}) \, (
k_{\0}^{\, \2} + \rho_{\0}^{\,\2})^{\2}}~~~~~\mbox{and} ~~~~~
\gamma = \mbox{$\frac{1}{2}$} \,
\frac{k_{\0}^{\2}+\rho_{\0}^{\2}}{k_{\0}^{\, \2} -
\rho_{\0}^{\,\2}} \, \epsilon \, \, .
\]
By convention we choose $\gamma$ to be positive. The linear term
in $\Delta \sigma $ with coefficient $\epsilon$ can be eliminated
by a shift in variables yielding a Breit-Wigner form for the
$\sigma$ (and hence energy) spectrum. However, we shall not do
this here since we intend to generalize the above formula to
fourth order in the next Section, and then it is not in general
possible to eliminate by a shift both of the odd powers in $\Delta
\sigma$.

The quadratic expression in square brackets in Eq.(\ref{qbw}) has
two complex conjugate zeros which appear as poles in our integrand
for $P(t)$. These are at
\begin{eqnarray}
\label{sigma}
 \sigma &= & \sigma_{\0} + \frac{k_{\0}^{\2}-
\rho_{\0}^{\2}}{k_{\0}^{\, \2} + \rho_{\0}^{\,\2}} \, \gamma \,
\pm \, i \, \frac{2\,k_{\0}\rho_{\0}}{k_{\0}^{\, \2} +
\rho_{\0}^{\,\2}}\,
\gamma \nonumber \\
 & \equiv &x_{\0} \pm \, i \, y_{\0} \, \, ,
 \end{eqnarray}
 with $x_{\0}$ and $y_{\0}$ the real and imaginary parts of
 $\sigma$ respectively.
The integral in Eq.(\ref{int}) is thus proportional to
\[\int_{\0}^{\infp} \hspace*{-.4cm} \mbox{d}\sigma \,
\exp (-i \, \sigma^{\2} \tilde{t}\,) \, \left[ \, (\sigma -
z_{\0})(\sigma - z_{\0}^*) \, \right]^{- \1} \, \, ,
\]
with $z_{\0}=x_{\0}+iy_{\0}$. Since by assumption this integral is
sharply peaked about $\sigma_{\0}$ we can formally extend the
lower limit of the integral from $0$ to $- \infty$. This is the
third approximation in this derivation. The theorem of residues
then yields
\[\int_{\infm}^{\infp} \hspace*{-.4cm} \mbox{d}z\,
\exp(-i \,z^{\2} \tilde{t} \, ) \,  \left[ \, (z - z_{\0})(z -
z_{\0}^*) \, \right]^{- \1} = (\pi / y_{\0}) \, \exp(-i
\,z_{\0}^{*\,\2} \tilde{t}\,) \, \, .
\]
Taking the square of the modulus of this we finally obtain, after
some simple algebraic manipulations,
\begin{equation}
\label{secord}
P(t) = \exp \left(- \frac{2 \,x_{\0} y_{\0}}{m
a^{\2}} \, t \right) \, \, .
\end{equation}
The above expression is of the desired exponential form
\begin{equation}
P(t) = \exp \left( - t/\tau_{\0}\right)\, \, ,
\end{equation}
normalized (automatically) to $P(0)=1$, and with predicted
lifetime
\begin{equation}
\label{tau0} \tau_{\0} = \frac{m a^{\2}}{2 \,x_{\0} y_{\0}}\, \, .
\end{equation}
The narrow width of the Breit-Wigner assumed in the derivation is
guaranteed if $\gamma \ll 1$ which corresponds to $\exp(2\, a
\rho_{\0} \, w) \gg 1$. This is just the condition anticipated in
the previous Section. Thus, for the validity of the above
derivation the choices of $E_{\0}$, $V_{\0}$, and $b-a$ must be
such as to satisfy this condition. In all our subsequent numerical
calculations this is indeed the case.

\section*{IV. HIGHER ORDER CORRECTIONS}

The calculation in the previous Section is a non-relativistic
quantum mechanical derivation of the exponential law. It is
incompatible with the existence of an experimental QZE and if
repeated for negative (as well as positive) values of $t$ it
yields the non-analytic expression,
\[ P(t) = \exp[-|t| / \tau ]\, \, . \]
This satisfies the symmetry property $P(-t)=P(t)$ (time reversal
invariance) consistent with the absence of all odd powers of $t$
(Section I), but it cannot be expressed as a power series in $t$.

What is  surprising, at first, is the following: If $P(t)$ is
calculated by evaluating the integral in Eq.(\ref{int}) {\em
numerically} for various $t$, and then, for small $t$, fit with a
polynomial, we find the odd powers of $t$ greatly suppressed. That
is, the numerical calculation is far nearer to what one may have
expected, a priori, on the basis of the arguments in our
introduction. This $P_{\mbox{\tiny num}}(t)$ is practically an
even function in $t$, and in particular the linear term in $t$ is
almost zero. To understand this we recall the expression of the
non-decay amplitude, see Eq.(\ref{pge}),
\[
P(t) = \left| \, \int_{\0}^{\infp} \hspace*{-.4cm} \mbox{d}E \, \,
|g_{\0}(E)|^{\2} \, e^{-iEt} \, \right|^{\2} \approx \left| \,
\int_{\infm}^{\infp} \hspace*{-.4cm} \mbox{d}E \, \,
|g_{\0}(E)|^{\2} \, \left[ \, \cos(Et) + i \, \sin (Et) \, \right]
\, \right|^{\2} \, \, .
\]
the terms in square bracket are bounded and by assumptions the
$|g_{\0}(E)|^{\2}$ is highly peaked around $E=E_{\0}$ and tends to
zero at least like $(E-E_{\0})^{- \2}$ when $E\to \pm \infty$. The
odd terms in $t$ lie in the sine term which is pure imaginary.
Hence, on taking the square of the modulus, all odd powers of $t$
disappear and our numerical calculation simply reflects this.
Indeed the presence of the very small odd terms in $t$ can simply
be attributed to rounding errors. Note that this result is derived
for a {\em Breit-Wigner spectrum}, which is consequently not in
itself a sufficient condition for obtaining the exponential law.
Thus our numerical integral of the Breit-Wigner would seem to be a
better approximation to the curve $P(t)$ than the analytically
derived exponential law. The following higher order approximation
confirms how difficult it is to derive a simple exponential
result.

We now generalize the calculation of the previous Section by
expanding $f(\sigma)$ up to fourth order in $\Delta \sigma$ . We
evaluate $f^{\mbox{\tiny $[4]$}}(\sigma)$ as an improved
approximation to the Breit-Wigner case. The rest of the procedure
remains the same. Up to second order we obtained
Eq.(\ref{secord}).
Now, we must evaluate
\[
\int_{\infm}^{\infp} \hspace*{-.4cm} \mbox{d}z\, \exp(-i \, z^{\2}
\tilde{t}\,) \, \left[ \, (z - z_{\1})(z - z_{\1}^*) (z -
z_{\2})(z - z_{\2}^*) \, \right]^{- \1} \, \, ,
\]
where in our subsequent plots the values of $z_{\1,\2}$, with real
and imaginary parts $x_{\1,\2}$ and $y_{\1,\2}$respectively, will
be obtained numerically. However, even without these calculations
we expect, from the second order approximation, that one of the
poles must correspond to $z_{\0}$. Let this be $z_{\1}$
($z_{\1}=z_{\0}$). The other then provides a deviation from the
exponential law (small if $y_{\2}\gg y_{\1}$). Indeed analytically
\[
\int_{\infm}^{\infp} \hspace*{-.4cm} \mbox{d}z\, \frac{\exp(-i\,
z^{\2} \tilde{t}\,)}{ (z - z_{\1})(z - z_{\1}^*) (z - z_{\2})(z -
z_{\2}^*)} = \frac{ \pi}{z_{\1}^* - z_{\2}^*} \, \left\{ \,
\frac{\exp( - i\,z_{\1}^{*\,\2} \tilde{t}\,)}{y_{\1} (z_{\1}^* -
z_{\2})} + \frac{ \exp( - i\,z_{\2}^{*\,\2} \tilde{t}\,)}{ y_{\2}
(z_{\1} - z_{\2}^*)} \, \right\}  \, \, .
\]
Defining
\[
\exp(i\,\alpha) = (z_{\1}^* - z_{\2}) / (z_{\1} - z_{\2}^*)
~~~~~\mbox{and}~~~~~\beta=x_{\1}^{\2} - x_{\2}^{\2} + y_{\2}^{\2}
-y_{\1}^{\2} \, \, ,
\]
we find
\begin{equation}
\label{p4} \hspace*{-.3cm} P^{\mbox{\tiny $[4]$}}(t) = N \,
\left\{ \, \exp(-4\,x_{\1}y_{\1} \tilde{t} \, ) +
\frac{y_{\1}^{\2}}{y_{\2}^{\2}} \, \exp(-4\,x_{\2}y_{\2} \tilde{t}
\, )  + 2 \, \frac{y_{\1}}{y_{\2}} \, \exp[-2\, ( x_{\1}y_{\1} +
x_{\2}y_{\2}) \, \tilde{t} \, ]   \, \cos (\alpha + \beta \,
\tilde{t} \, ) \, \right\}\, \, ,
\end{equation}
where the normalization at $t=0$, $P(0)=1$, fixes
\[ N = \left[ \, 1
+ \frac{y_{\1}^{\2}}{y_{\2}^{\2}} + 2 \, \frac{y_{\1}}{y_{\2}} \,
\cos \alpha \, \right]^{- \1}\, \, .
\]
From Eq.(\ref{p4}) we see that as $y_{\1}/y_{\2} \rightarrow 0$
the single exponential is indeed recovered. This justifies  our
prediction that $z_{\1}=z_{\0}$, in agreement also with our
numerical derivations of $z_{\1}$. Apart from the coefficients in
front of the additional terms in $P^{\mbox{\tiny $[4]$}}(t)$ we
observe that for $ x_{\2}y_{\2} \gg x_{\1}y_{\1}$ (valid in all
our numerical solutions for $z_{\1}$ and $z_{\2}$) the
exponentials in these terms fall off far more rapidly than the
leading exponential in $t$. Thus, they are of interest only for
small $t$.

The surprise here (at least for the authors) is that
$P^{\mbox{\tiny $[4]$}}(t)$ {\em displays exactly the absence of a
linear term}, notwithstanding that it is still an approximation,
only a step above the Breit-Wigner approximation. This is
independent of the specific expressions for $z_{\1,\2}$ which
would be far too complicated to be given algebraically in terms of
the parameters. To see this, we expand $P^{\mbox{\tiny $[4]$}}(t)$
as a power series in $\tilde{t}$. The coefficient of the linear
term $\tilde{t}$
after a little algebra can  be shown to be {\em zero
identically}. To avoid any misunderstandings, we emphasize that
this fact has nothing to do with numerical integrations. We have
used the theorem of residues to evaluate the integrals in the
above. Anticipating one of our conclusions, one therefore sees
that one must make a very precise set of approximations to derive
the exponential law.

\section*{V. NUMERICAL POLE CALCULATIONS AND INVERSE ZENO EFFECT}

So far we have not derived any numerical values for the poles. In
this Section, we shall do so with the principle objective of
finding a phenomenological fit to one of the time constants that
characterizes the improvement to the exponential curve. In the
expression for $P^{\mbox{\tiny $[4]$}}(t)$ three parameters appear
$x_{\1}y_{\1}$, $x_{\2}y_{\2}$, and $\beta$ . The lifetime $\tau$
is no longer sufficient by itself to parameterize $P(t)$. One
would like to interpret these constants, or more precisely, the
related time parameters, in physical terms.

In general the pole positions are complicated functions of the
input parameters $a$, $b$, $V_{\0}$ and $E_{\0}$ of the model. The
only one we give here is an approximate algebraic expression for
the $P^{\mbox{\tiny $[2]$}}(t)$ lifetime ($\tau_{\0}$) derived in
Section III. By using Eqs.(\ref{sigma}) and (\ref{tau0}), we find
\begin{equation}
\tau_{\0} = m a^{\2}  \left[ \,   \frac{k_{\0}
\rho_{\0}}{k_{\0}^{\, \2} - \rho_{\0}^{\,\2}} \, \left(\, 2 \,
\sigma_{\0} \, \epsilon + \epsilon^{\2} \, \right) \, \right]^{-
\1} \, \, .
\end{equation}
Dropping the term proportional to $\epsilon^{\2}$,
\begin{equation}
\tau_{\0} \approx \frac{V_{\0}^{^{\2}}}{16 \, \left[ \,
E_{\0}\,(V_{\0}-E_{\0})\, \right]^{ \mbox{\tiny $3/2$} }} \,
\left[ \, 1+a\sqrt{2 \, m (V_{\0}-E_{\0})} \, \right]\,\exp \left[
\, 2 \,(b-a)\sqrt{2 \, m (V_{\0}-E_{\0})}\, \right ]\, \, .
\end{equation}
To study the other time parameters it is far easier to select the
input values and solve numerically the equation for the pole
positions,
\begin{equation}
\label{f4} f^{\mbox{\tiny $[4]$}}(\sigma) = 0 \, \, . \
\end{equation}
It is convenient, as far as possible, to work with  dimensionless
quantities. In Table 1, we show for various values of $V_{\0}$ the
solution of the above equation for $z_{\1}$ and $z_{\2}$. The
multiple of $\pi$ (a convenient choice suggested by the
transcendental equation) that appears in the header of each
sub-table corresponds to the number of quasi-stationary energies.
For each value of these energies we list, for selected $w$, the
values of $x_{\1,\2}$, $y_{\1,\2}$, and the dimensionless
parameters $\tau_{\1,\2}/ma^{\2}$ defined below. The values of
$z_{\1}$ coincide within numerical accuracy with those of
$z_{\0}$, the pole position for $f^{\mbox{\tiny $[2]$}}(\sigma) =
0$. Consequently, to a good approximation, this pole still
determines the lifetime for the process
\begin{equation}
\tau \approx \tau_{\1} = \frac{m \, a^{\2}}{2\, x_{\1} y_{\1}} \,
\, .
\end{equation}
The other pole produces the modification from the simple
exponential curve and determines a second exponential time
parameter $\tau_{\2} \ll \tau_{\1}$ ($x_{\1}y_{\1} \ll
x_{\2}y_{\2}$)
\begin{equation}
\tau_{\2} = \frac{m \, a^{\2}}{  2\, x_{\2} y_{\2}} \, \, .
\end{equation}
In Fig.\,1 we display the dependence of $\tau_{\2}$ upon $w$ for
various $E_{\0}$ and an arbitrary chosen value of $V_{\0}$. In
Fig.\,2 we show the dependence of $\tau_{\2}$ upon
$\sqrt{2\,m(V_{\0}-E_{\0})}$ for different values of $w$. From the
above curves we see that, to a very good approximation,
$\tau_{\2}$ grows linearly with $w$ and hence with the barrier
width $b-a$. It is also proportional approximately to $[\,2 \,m
(V_{\0}-E{\0})\,]^{\mbox{\tiny $-1/2$}}$. This for the parameter
ranges considered. We defer an interpretation of these results to
our conclusions.

For $t\rightarrow 0$, the exponential survival law must
necessarily lie below the physical curve $P(t)$ if this lacks a
linear term since both are normalized to $P(0)=1$.
The fact that the lifetime measures the mean time for decay means
that an exponential curve with \emph{said lifetime as input} must
eventually rise above $P(t)$ to compensate for the small time
"short-fall", i.e. there is a cross-over point.
As an aside, we note that the power law in $t$ predicted for
$P(t)$ at long times means that any exponentially falling curve
must lie below $P(t)$ asymptotically adding another cross-over
point. The first cross-over allows for  the so-called {\em inverse
Zeno effect} (IZE)\cite{ize}. This says that if a first time
measurement is made after this cross-over point (but logically
before any second cross-over point) the system will have been
found to decay at a rate greater than "expected".

First, let us analyze critically this definition. Unlike for the
QZE this measurement or these measurements (if repeated at the
appropriate time intervals) after the inverse Zeno cross-over only
contradicts our {\em expectations} based upon the use of an
exponential curve with the experimental lifetime. But the choice
of exponential is somewhat arbitrary. We can give at least two
alternative proposals for the reference exponential curve. The
first, which is the most natural phenomenologically, is to refer
to an exponential best fit to $P(t)$. This fit almost certainly
will \emph{not} have the exact same lifetime as $P(t)$. The second
possibility is that one may compare $P(t)$ with a {\em
theoretical} single exponential (possibly from a model calculation
as in our case). These possibilities distinguish themselves from
the original because they do not {\em necessarily} have a
cross-over point and hence need not imply the conditions for an
IZE.

We have a natural choice of exponential in our model, the
Breit-Wigner exponential. We have thus looked for cross-over
points by confronting $P^{\mbox{\tiny $[4]$}}(t)$ with
$P^{\mbox{\tiny $[2]$}}(t)$. In all cases examined they have been
found.  Thus, even with our choice of
 exponential the IZE  is possible.
We can see this  more clearly by
 observing that, in our tables, $y_{\1}/y_{\2} \ll 1$ and $x_{\1} \approx
x_{\2}$, thus we can approximate Eq.\,(\ref{p4}) as follows
\begin{equation}
\label{p4bis} P^{\mbox{\tiny $[4]$}}(t) \approx \exp(-4\,
x_{\1}y_{\1} \tilde{t} \, ) \, \left\{ \, 1 + 2 \,
\frac{y_{\1}}{y_{\2}} \, \left[ \, \exp(-2  \, x_{\2} y_{\2}
\tilde{t}\,)   \, \cos (\alpha + \beta \, \tilde{t} \,)  - \cos
\alpha \right] \, \right\}\, \, .
\end{equation}
Consequently,
\begin{equation}
\label{p4p2} \frac{P^{\mbox{\tiny $[4]$}}(t)}{P^{\mbox{\tiny
$[2]$}}(t)} \approx 1 + 2 \, \frac{y_{\1}}{y_{\2}} \, \varphi(t)
\, \, ,
\end{equation}
where
\[
\varphi(t) = \exp \left(-2 \, \frac{x_{\2} y_{\2}}{ma^{\2}} \, t
\right) \, \cos \left( \alpha + \frac{\beta}{ma^{\2}} \, t \right)
- \cos \alpha \, \, .
\]
Intersection points occur when $\varphi(t)=0$, i.e. when
\begin{equation}
\label{izpoint} \exp \left(2 \, \frac{x_{\2} y_{\2}}{ma^{\2}} \, t
\right) = \cos \left( \alpha + \frac{\beta}{ma^{\2}} \, t \right)
/ \cos \alpha \, \, .
\end{equation}
The trivial solution is, of course, $t=0$. The condition for  the
IZE can be determined from the relative values of the time
derivative of the functions on the left and right hand sides of
the above equation at $t=0$.  The necessary and sufficient
condition for a non-trivial intersection is
\begin{equation}
\label{crossover}
 2 \, x_{\2} y_{\2} < - \, \beta \,\tan\alpha\, \,  .
\end{equation}
This is because, with this condition satisfied, the exponential
starts off with a flatter growth than the oscillating and bounded
right hand side of Eq.(\ref{izpoint}), and hence they must
necessarily cross.

If  $x_{\1}=x_{\2}$ ($\Rightarrow \sin \alpha =0$), we find only
the trivial solution $t=0$. On the other hand, from an examination
of Table 1, we see that
\[- \, \beta \, \tan\alpha = 2 \, \frac{\left(x_{\1}^{\2} - x_{\2}^{\2} + y_{\2}^{\2}
-y_{\1}^{\2}\right)
 \left( x_{\1}- x_{\2}\right)\left( y_{\1}+ y_{\2}\right)}
{\left( x_{\1}- x_{\2}\right)^{\2} - \left( y_{\1}+
y_{\2}\right)^{\2}} \approx 2 \, x_{\2} y_{\2} \,  \left[ 1 -
\left(\frac{y_{\2}}{x_{\1}-x_{\2}}\right)^{\2} \, \right]^{- \1}
\gg 2 \,x_{\2} y_{\2}\, \, .
\]
Hence a cross-over  occurs for all our tabulated cases. In Fig.\,3
we show examples of this by plotting $\varphi(t)$ against $t$ for
several quasi-stationary states $E_{\0}$ of a fixed barrier height
$V_{\0}$ and an arbitrary chosen value of $w$. The large dots
identify the cross-over times.

\section*{VI. CONCLUSIONS}

We have discussed in this paper the survival law and its relevance
to the QZE and the IZE within the context of a simple potential
model. We have shown that the general arguments that allow for the
QZE based upon the hermiticity of the Hamiltonian are indeed valid
in this model unless one makes very specific assumptions and/or
approximations. In particular, the exponential law, which does not
allow for  the QZE, is obtained only if two approximations are
made:

 1 -
  The denominator $f(\sigma)$ in the
spectral integral for the amplitude of non-decay is so highly
peaked about the initial quasi-state that only the lowest - second
order - terms in $\Delta \sigma$ need be considered;

 2 -
The integral in energy $E$ or equivalently $\sigma$ is extended
below the threshold to $- \infty$.\\
Point 2 is made in order to apply the theorem of residues
essential for deriving a single potential survival law. It is also
the cause of loss of analyticity in $t$, although this is normally
ignored by arguing that only positive values of $t$ are physically
significant.

Our analysis has gone on to show that while point 1 (the
Breit-Wigner spectrum) is not the sole cause for the theoretical
lose of the QZE, it is the primary one. Indeed, if one includes
higher order corrections, such as the fourth order terms in
$\Delta \sigma$, then the resultant $P^{\mbox{\tiny $[4]$}}(t)$
displays the absence of linear term in $t$, however small the
extra contributions might be.

This result was somewhat of a surprise. One could have reasonably
supposed that only in the limit of an exact calculation, i.e. when
all orders of $\Delta \sigma$ are included, is the linear term
absent. Indeed the authors only expected that to fourth order the
coefficient of the linear term would be reduced when compared to
that of the exponential curve. Instead we have been able to show
that it is {\em rigorously null} already in $P^{\mbox{\tiny
$[4]$}}(t)$. Note that to prove this we have applied points 2 and
3. Point 2 is also significant for another reason. It is known to
be the cause for the lose of the long-time power law behavior, and
this is independent of any other approximation made. Our
conclusions upon the question of the theoretical prerequisite for
the QZE is that it is indeed a feature of $P(t)$ and that one must
make very specific assumptions or approximations to avoid this
effect. Only with the explicit exclusion of short and long times,
can a single exponential curve be a good approximation to $P(t)$.

We have also considered in the previous Section the question of
the existence or otherwise of the IZE. There is in our opinion a
measure of ambiguity in the condition for the existence  of this
effect because of a non unique choice of the exponential used as
reference. Since any exponential is at best an approximation to
the physical curve, its definition is subject to discussion. An
exponential with the same lifetime as the real curve is the
conventional choice. However, there are other choices for which
the possible  existence of an IZE is far from obvious. We have
shown that one possibility, the comparison of $P^{\mbox{\tiny
$[4]$}}(t)$ with $P^{\mbox{\tiny $[2]$}}(t)$, does indeed allow
for  an IZE. Nevertheless, we do not consider the IZE and the QZE
comparable phenomena. The former (IZE) refers to our
"expectations", and as we have argued our expectations are subject
to some ambiguity. The latter (QZE) predicts a physical
suppression of decay through continuous observation and it is
completely independent of the existence or otherwise of an
approximate exponential curve.

Of particular interest in our results is the dependence of the
second exponential time parameter $\tau_{\2}$. The first time
parameter $\tau_{\1}$ is essentially the lifetime of the system
and has a characteristic exponential rise with barrier width
$b-a$, whereas this second time parameter can be phenomenological
represented by
\begin{equation}
\tau_{\2} \approx \frac{ m a^{\2} \, w}{
\sqrt{2\,ma^{\2}(V_{\0}-E_{\0})}}= \frac{ m \, (b-a)}{
\sqrt{2\,m(V_{\0}-E_{\0})}} \, \, .
\end{equation}
This form is very suggestive. Consider an energy eigenstate {\em
above the barrier}, $E>V_{\0}$. In the  barrier region $(a<r<b)$
the particle will have velocity $v = \sqrt{2\,m(E-V_{\0})}/ m$.
Consequently the time taken in crossing the barrier will be:
\begin{equation}
        \Delta t = \frac{m (b-a)}{\sqrt{2\,m(E-V_{\0})}}.
\end{equation}
This "mirrors" the expression for $\tau_{\2}$ (except for the
feature that $E_{\0}$ has a discrete spectrum). Both tend to
infinity if the particle energy and potential energy are set
equal. A type of consistency condition. Furthermore, both grow
linearly with barrier width. This tempts us to speculate (and at
this stage it is no more than an hypothesis) that $\tau_{\2}$ is a
measure of the time that the particle takes to tunnel free.
Equivalently, we may say that $\tau_{\2}$ is the time during which
the particle is neither bound nor free\cite{timelapse}. However,
since this infringes upon the question of transit times and
super-luminal velocities \cite{SL}, which is a very topical, but
also complex subject, we desist from any further considerations at
this point. The tau are not the only time parameters in
$P^{\mbox{\tiny $[4]$}}(t)$, we also have 1/$\beta$ in the
oscillatory term. Mathematically it is the natural consequence of
interference between the two exponentials of the non-decay
amplitude. It is therefore a wave-like property of the particle.\\


\noindent {\bf \small ACKNOWLWDGEMENTS.} The authors wish to thank
Giampaolo C\`o and Piergiulio Tempesta for several comments during
the preparation of this manuscript and Saverio Pascazio for
helpful suggestions upon the revised version. One of the authors
(SDL) also gratefully acknowledges the FAEP (University of
Campinas), INFN (Theoretical Group) and MIUR (Department of
Physics) for financial support during his stay at the University
of Lecce where this paper was prepared.

\newpage

\begin{table}
\hspace*{-5.49cm}
\includegraphics[width=23.28cm, height=18.cm, angle=0]{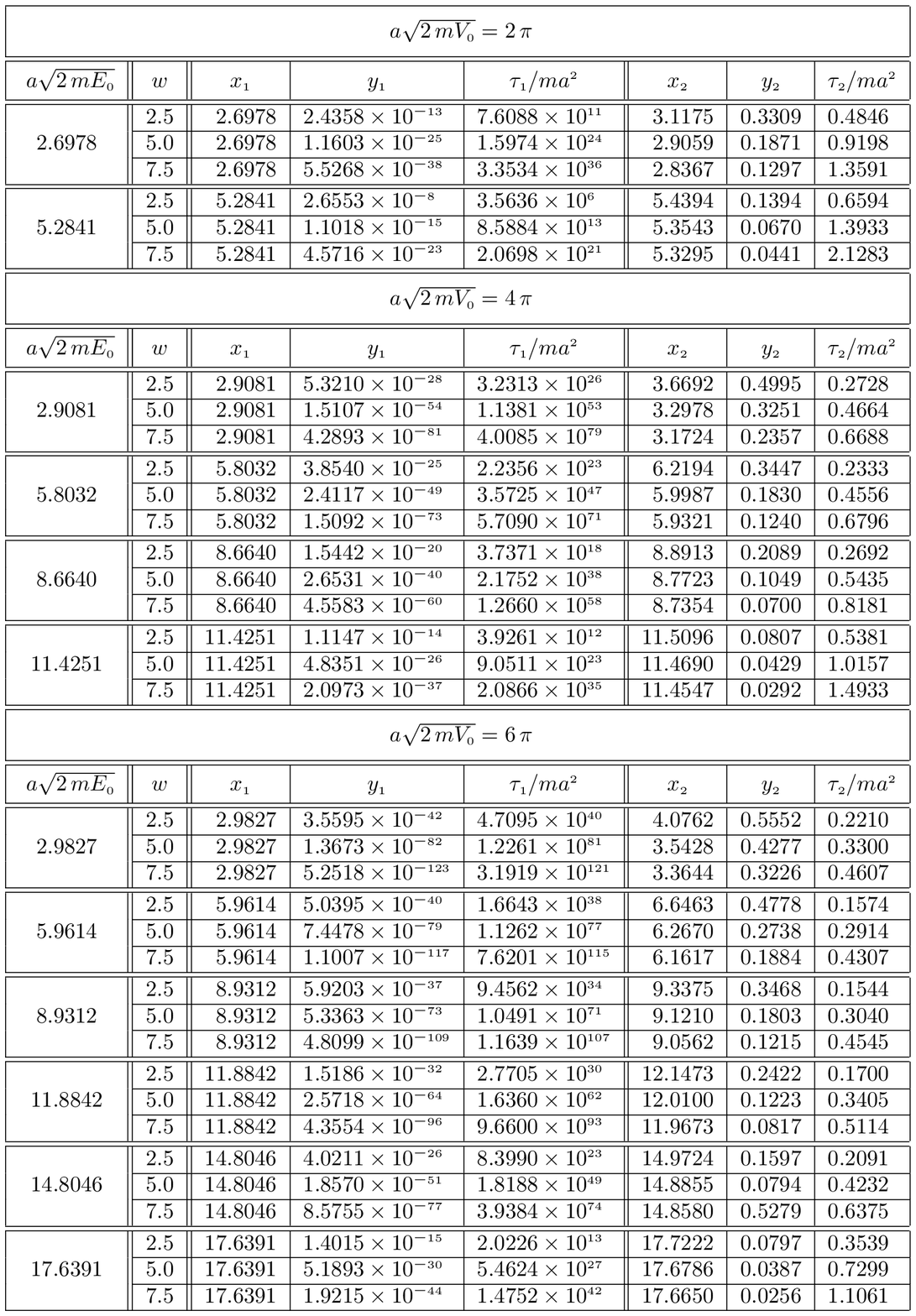}
\hspace*{-4.1cm}
\includegraphics[width=20.5cm, height=15.55cm, angle=0]{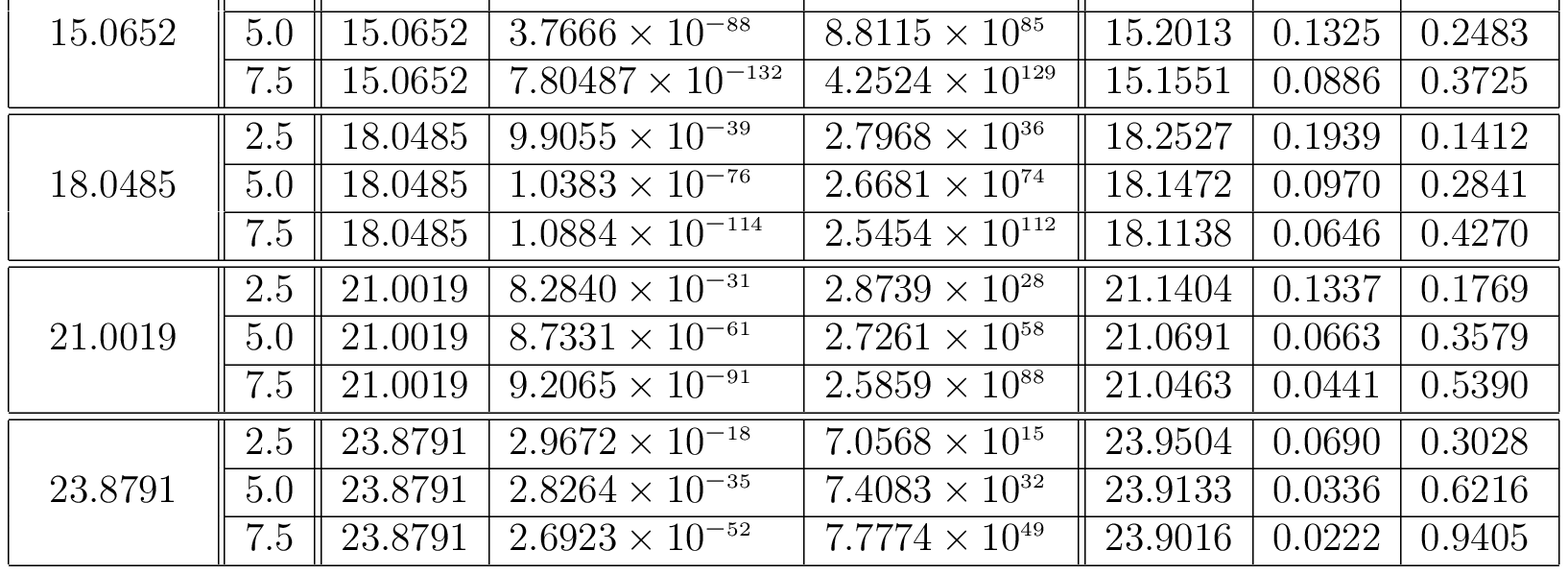}
{\vspace*{-12cm}\caption{Numerical values of $x_{\1,\2}$\,,
$y_{\1,\2}$ and $\tau_{\1,\2}$ for chosen values of $V_{\0}$. Each
block lists the parameters for all the quasi-stationary
eigenvalues $E_{\0}$.}}
\end{table}

\newpage

\begin{figure}[hbp]
\begin{center}
\hspace*{-2.5cm}
\includegraphics[width=20cm, height=22cm, angle=0]{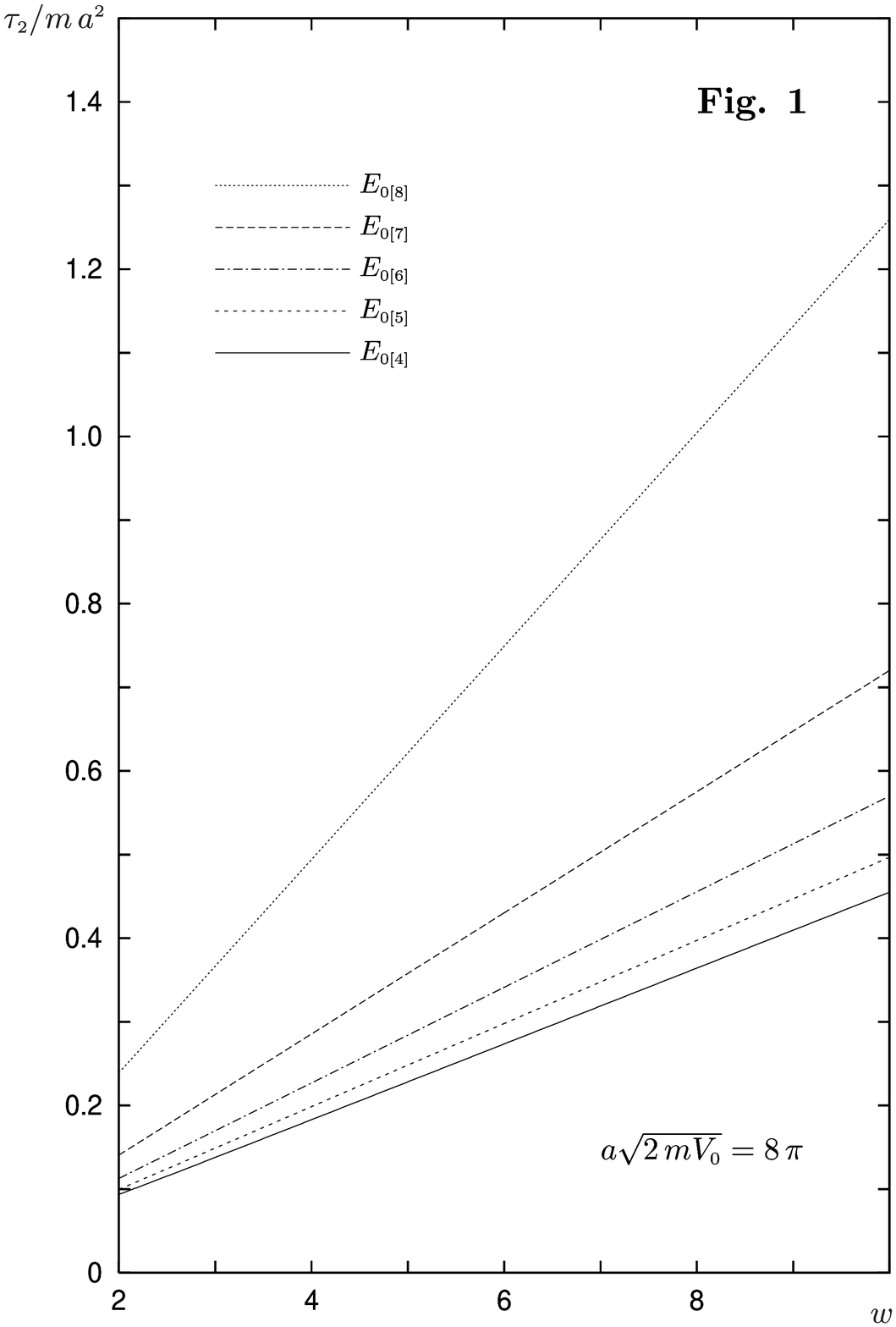}
\caption{Dependence of $\tau_{\2}$ on $w$ for several of the
quasi-stationary energies $E_{\0}$ of a fixed $V_{\0}$}
\end{center}
\end{figure}

\newpage

\begin{figure}[hbp]
\hspace*{-2.5cm}
\includegraphics[width=20cm, height=22cm, angle=0]{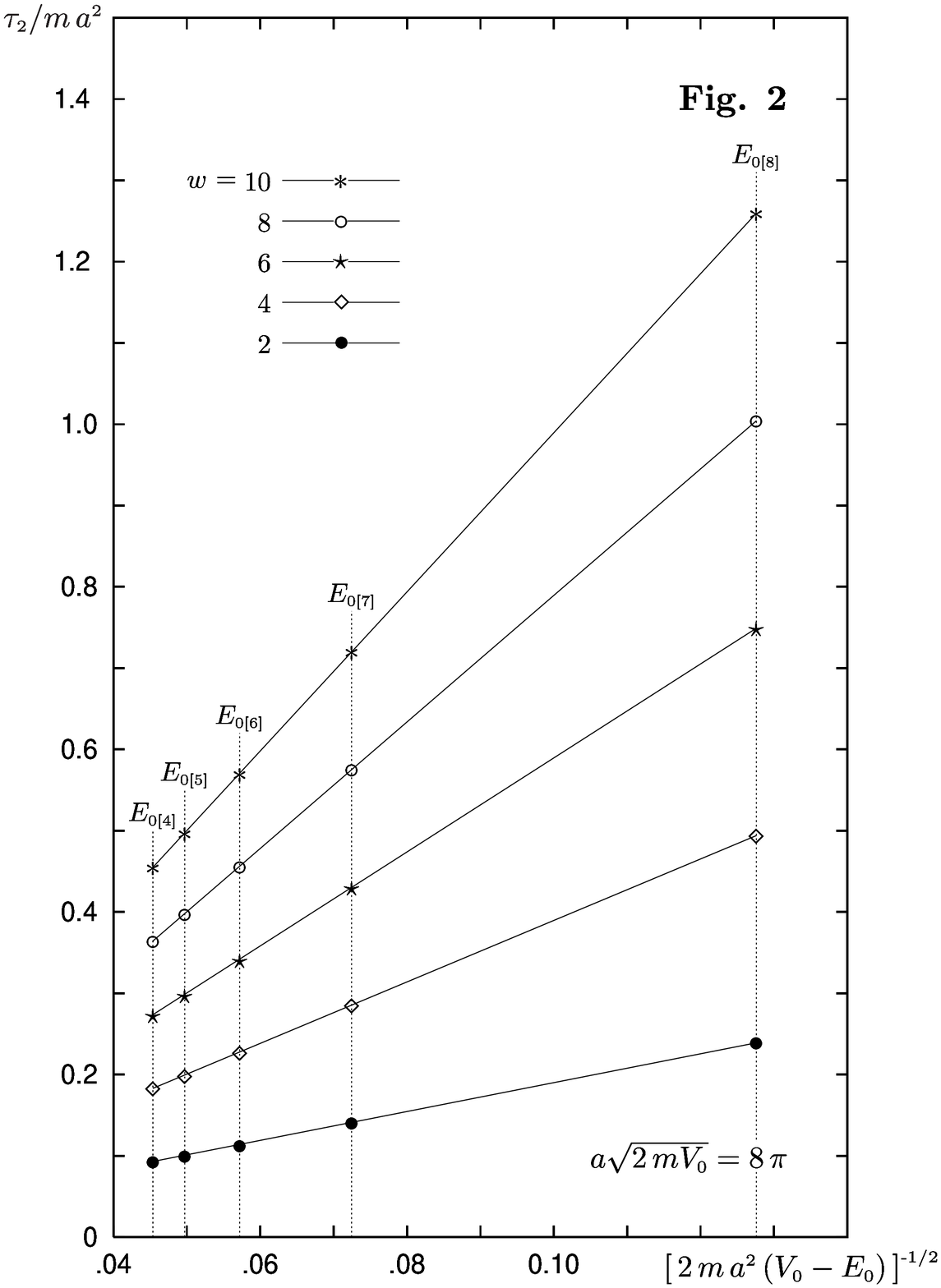}
\caption{$\tau_{\2}$ vs  1/$\sqrt{V_{\0}-E_{\0}}$ for various
quasi-energies $E_{\0}$ and $w$ values. The lines drawn
demonstrate excellent linear fits, for a fixed $w$.}
\end{figure}

\newpage

\begin{figure}[hbp]
\hspace*{-2.5cm}
\includegraphics[width=20cm, height=22cm, angle=0]{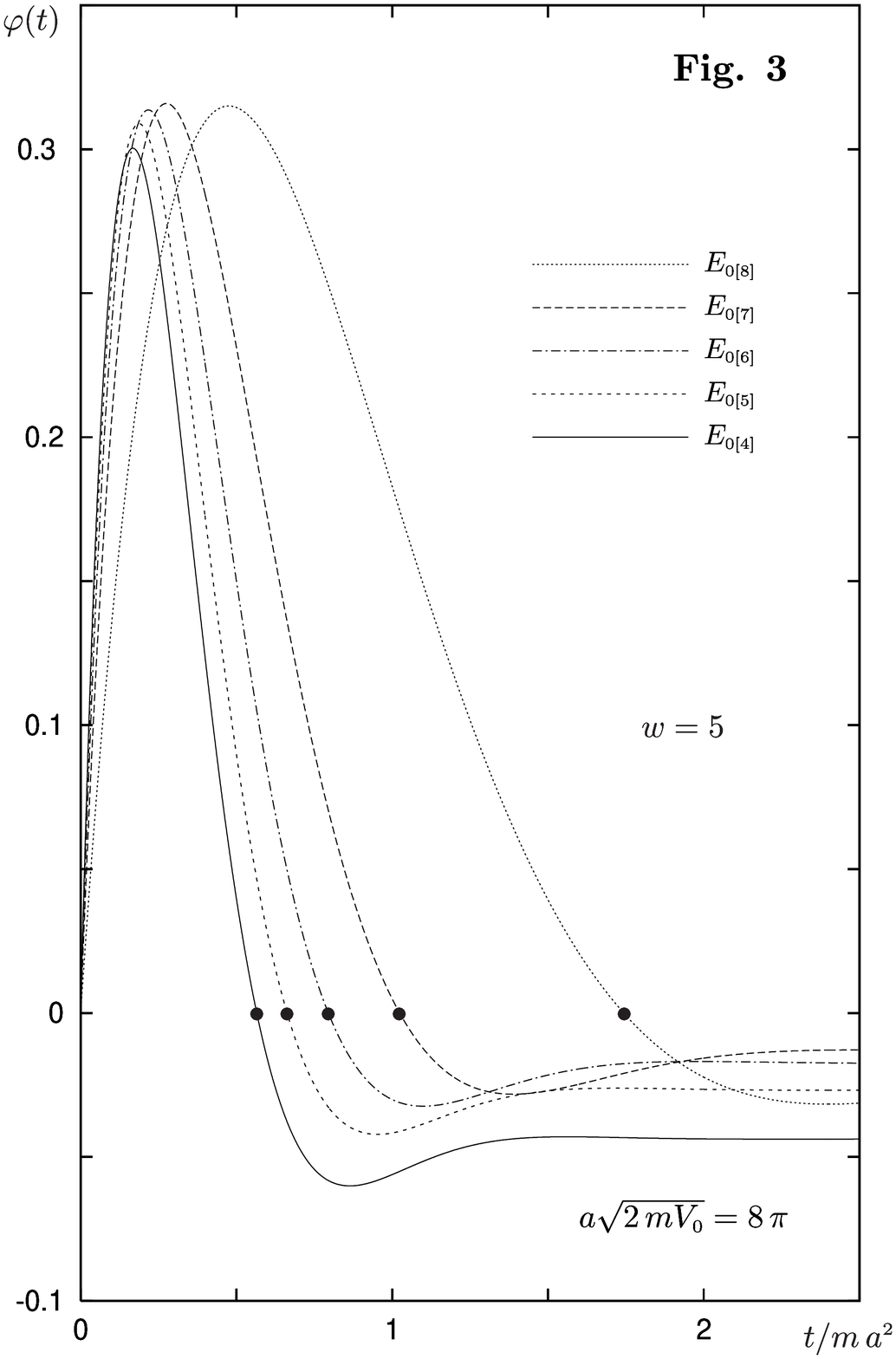}
\caption{The function $\varphi(t)$ vs $t$ for the indicated values
of $V_{\0}$ and $\omega$. The five curves are for different values
of $E_{\0}$ . For each of the curves the cross-over points are
indicated by the large dots. }
\end{figure}

\end{document}